# Time Variability of Fibrillatory Waves Energy Predicts Long-Term Outcome of Atrial Fibrillation Concomitant Surgical Ablation

Juan Ródenas[1], Pilar Escribano[1], Miguel Martínez-Iniesta[1], Manuel García[1], Fernando Hornero[2], José J Rieta[3], Raúl Alcaraz[1]

[1] Research Group in Electronic, Biomedical and Telecommunications Engineering, University of Castilla-La Mancha, Albacete, Spain
[2] Cardiovascular Surgery Department, Hospital Clínico Universitario de Valencia, Spain
[3] BioMIT.org, Electronic Engineering Department, Universitat Politecnica de Valencia, Spain

**Abstract**

*Surgical ablation (SA) is the most effective procedure to terminate atrial fibrillation (AF) in patients requiring concomitant open-heart surgery. However, considering the great stress provoked in the patient's heart, along with the benefits of anticipating antiarrhythmic therapeutical decisions, preoperative prediction of long-term failure of the procedure is an interesting clinical challenge. Hence, the present work introduces a novel algorithm to anticipate SA outcome after one year of follow-up by just analyzing the surface ECG. The method firstly extracts fibrillatory waves reflected on standard lead V1 using an adaptive QRST cancellation approach. The resulting signal is then segmented into 1 s-length intervals and wavelet energy is computed for all of them. Finally, the coefficient of variation of the time series obtained for the 7th scale is computed. Analyzing 20 second-length preoperative ECG excerpts from 53 persistent AF patients undergoing concomitant open-heart surgery, only the proposed method reported statistically significant differences between the patients who relapsed to AF and those who maintained sinus rhythm during the follow-up. The algorithm also provided values of sensitivity, specificity, and accuracy between 10 and 20% better than the well-established dominant atrial frequency and fibrillatory waves amplitude, thus suggesting to be a promising predictor of AF recurrence after SA.*

## 1. Introduction

Atrial fibrillation (AF) is the most common cardiac arrhythmia in clinical practice, affecting more than 33 million people worldwide, i.e., more than 0.5% of the world population [1]. Its prevalence is directly related to age. Whereas only 0.16% of the population under 50 years suffer from this disease, this percentage grows to 17% within people over 80 years old [2]. Hence, current projections about the growth of the population over 60 years old, from 962 million in 2017 to over 2,000 million in 2050, the increased prevalence of chronic diseases predisposing to AF in older people, and continuous improvements in AF detection suggest that the incidence of this arrhythmia will significantly increase in coming years [3]. This situation has an huge economic impact on every health system, because it is estimated that AF management represents approximately 1% of the world health budget and more than 15% of the expenditure for cardiovascular diseases [4].

Unfortunately, the mechanisms triggering and supporting AF are still not completely known and the available treatments are not as effective as desired from a clinical point of view [5]. Indeed, although current therapies, including pharmacological cardioversion, electrical cardioversion, and catheter ablation, present initial success rates higher than 90%, AF recurrence is common after several weeks or months [3]. Contrarily, surgical ablation (SA) has reported a notably higher long-term success rate about 80% [6]. This procedure consists of an open-heart surgery, where ablation lines are created in the atria to redirect sinus node impulses to the atrioventricular node through a specific route [7]. The intervention can be applied independently or concomitantly with other cardiac procedures, including coronary artery bypass grafting, or reparation of mitral and aortic valves [8].

After SA, patients are commonly treated with drugs, and routinely evaluated at 3, 6, 12 months, and then every year. If a stable sinus rhythm (SR) is maintained for several months, the drugs are removed. However, in the case of AF recurrence, electrical cardioversion is applied to restore SR [6]. Given this context, a preoperative prediction of patient's rhythm at one year post-surgery could be very useful to plan a tailored follow-up of each patient, thus scheduling in advance electrical cardioversion for those patients with a low probability of SR maintenance and avoiding aggressive drug treatments in the remaining patients [9].

Although some clinical parameters, such as age, left atrial size or preoperative time in AF, among others, have provided a moderate ability to predict long-term outcome of SA [10], other indices computed from the surface electrocardiogram (ECG) recording have shown great effectiveness to anticipate immediate outcome of the procedure at discharge [11]. Hence, the present work introduces a new method to preoperatively predict long-term failure of SA based on estimating time variability of the wavelet energy contained by the fibrillatory ($f$-) waves reflected on the surface ECG signal.

## 2. Methods

### 2.1. Study population

A database composed of 20 second-length ECG segments extracted from standard lead V1 for 53 patients was analyzed in the present work. All the subjects were in persistent or permanent AF for, at least, four months before undergoing a open-heart surgery concomitantly with SA of AF. The ECG recordings were acquired less than 48 hours before the surgery with a sampling frequency of 1 kHz and an amplitude resolution of 0.4 $\mu$V. After one year of follow-up, 23 patients relapsed to AF (43.40%), whereas the remaining 30 maintained SR (56.60%).

### 2.2. Extraction of the $f$-waves

Only lead V1 was assessed in this study because it reflects the largest $f$-waves compared with the ventricular activity [12]. To improve further analysis and reduce different kinds of noises, this signal was initially preprocessed. Thus, a filtering methodology based on stationary wavelet transform (SWT) shrinking was firstly used to remove the powerline interference and simultaneously preserve original morphology of the $f$-waves [13]. Next, baseline wander was eliminated by making use of an IIR high-pass filtering with a cut-off frequency of 0.5 Hz [14]. Finally, high frequency noise was reduced through an IIR low-pass filtering with a cut-off frequency of 70 Hz [14]. Note that both filters were designed using a Chebyshev window with a relative sidelobe attenuation of 40 dB and applied in a forward/backward fashion.

Additionally, to reliably analyze the $f$-waves, they were firstly extracted from the preprocessed ECG signal by making use of a well-established QRST cancellation method [15]. Briefly, once all R-peaks were detected through an efficient phasor transform-based approach [16], QRST complexes were aligned. For every complex under cancellation, cross-correlation was used to select the 10 ones most similar. Next, a representative template of that complex was obtained using principal component analysis and finally subtracted from the preprocessed ECG.

### Time variability of the $f$-waves energy

The $f$-waves are characterized by a time-varying morphology, presenting waveforms of different sizes, amplitude and timings [17]. However, in the last years the degree of organization of these waves has been strongly correlated with a successful result in different treatments of AF, such as electrical cardioversion and catheter ablation [18]. Thus, in the present work a novel algorithm to estimate time course of the $f$-waves organization is proposed to anticipate long-term outcome of SA. More precisely, the signal containing the $f$-waves was segmented into 1 second-length intervals. This time period was obtained experimentally after several analyses. Next, every interval was decomposed into several time and frequency scales using a SWT. This transformation is an appropriate tool for the analysis of transients, aperiodicities and other non-stationary signal features, since subtle changes in the signal morphology can be easily highlighted over the scales of interest [19].

Having in mind that the $f$-waves present most relevant information in low frequency and that the ECG recordings were acquired with a sampling rate of 1 kHz, an eight-level wavelet decomposition was chosen. As in previous works [20], a sixth-order Daubechies function was used as mother wavelet. Then, the relative wavelet energy (RWE) for the 7-th scale was computed as

$$RWE7 = \frac{\sum_{k=1}^{N} C_j(k)^2}{\sum_{j=1}^{8} \sum_{k=1}^{N} C_j(k)^2}, \quad (1)$$

where $C_j$ is a vector containing the wavelet coefficients for the $j$-th scale and $N$ is the length of the original signal (i.e., the number of samples corresponding for a 1 second). Note that this scale was analyzed, because it covers the typical AF frequency range, i.e. 4–8 Hz [21]. Finally, the coefficient of variation of RWE7 (referred to as CVRWE7) for all ECG intervals was obtained as the ratio between its standard deviation and mean.

### 2.4. Performance Assessment

As a reference for comparison, two well-known features of the $f$-waves, such as dominant atrial frequency (DAF) and $f$-waves amplitude (FWA), were also computed. Thus, DAF was estimated as the frequency with the highest power spectral density amplitude within the 3–12 Hz range [21], and FWA was obtained as the root mean square of the $f$-waves [22].

The results for the three metrics, i.e. CVRWE7, DAF and FWA, were obtained as mean ± standard deviation. Moreover, statistical differences between the patients who relapsed to AF and maintained SR after the follow-up were tested by a Mann-Whitney $U$ test, a statistical sig-nificance $p < 0.05$ being considered as statistically significant. Also, the ability of each feature to discriminate between both groups of patients was evaluated by means of a receiver operating characteristic (ROC) curve. This plot is the result of plotting the fraction of true positives (TP) out of positives (sensitivity) against the fraction of false positives out of the negatives (1−specificity) at various threshold settings. Sensitivity (Se) was considered as the percentage of patients who relapsed to AF correctly classified. Similarly, the rate of patients maintaining SR was considered as the specificity (Sp). The optimal threshold was selected as the one providing the highest percentage of patients correctly classified, i.e., the largest accuracy (Acc). The area under the ROC curve (AROC) was also obtained as an aggregate measure of performance of a variable across all possible classification thresholds. Finally, values of positive ($P^+$) and negative predictivity ($P^-$) were computed as the proportion of truly identified patients relapsing to AF and maintaining SR, respectively.

Table 1. Mean and standard deviation values for the analyzed metrics from the patients who maintained SR and relapsed to AF. Statistical significance is also shown.

|  | Group of patients | | |
| --- | --- | --- | --- |
| Index | maintaining SR | relapsing to AF | $p$-value |
| CVRWE7 | 0.203±0.077 | 0.266±0.071 | 0.003 |
| DAF (Hz) | 6.10±1.40 | 6.70±0.90 | 0.060 |
| FWA ($\mu$V) | 48.70±17.80 | 44.70±22.70 | 0.201 |

## 3. Results

Mean and standard deviation values obtained for the three analyzed indices both from the patients who maintained SR and relapsed to AF during the follow-up are shown in Table 1. As can be seen, CVRWE7 was the only index that exhibited statistically significant differences ($p < 0.05$) between both groups of patients, reporting larger values for those who relapsed to AF than for those who maintained SR. Nonetheless, DAF provided a $p$-value very close to be significant, also the patients who maintained SR showing lower values than those relapsing to AF.

According to these findings, the best classification results were reached by CVRWE7, such as Table 2 displays. In fact, this metric obtained values of Acc and AROC between 16 and 23% higher than DAF and FWA. Similarly, improvements between 15 and 30% were also achieved by CVRWE7 in terms of Se, Sp and $P^+$, compared with the remaining parameters. Regarding $P^-$, DAF only reported a value 3% lower than CVRWE7, but that provided by FWA was 17% lower.

Table 2. Classification results obtained for the three analyzed metrics.

| Index | Se | Sp | Acc | AROC | $P^+$ | $P^-$ |
| --- | --- | --- | --- | --- | --- | --- |
| CVRWE7 | 65% | 80% | 74% | 80% | 78% | 70% |
| DAF | 52 | 63% | 58% | 58% | 48% | 67% |
| FWA | 46 | 59% | 53% | 57% | 52% | 53% |

## 4. Discussion

In previous works, DAF and FWA provided to be promising preoperative predictors of early AF recurrence after SA and before discharge [11]. According to these studies, in the present work lower and higher values of DAF and FWA were observed for the patients who maintained SR than for those who relapsed to AF. However, in this case no statistically significant differences were noticed between both groups. Moreover, whereas both indices reached classification rates higher than 70% at discharge [11], accuracy values lower than 60% were only achieved at the end of the first year of follow-up. Although several causes could contribute to these conflicting results, the most relevant one could be the atrial substrate alteration provoked by antiarrhythmic drugs, and even electrical cardioversion, during the follow-up. Nonetheless, this supposition requires further investigation.

In contrast to these two predictors, the proposed CVRWE7 reported statistically significant differences between both groups of patients. In fact, notably higher time variability in the $f$-waves energy was noticed for the patients who relapsed to AF than for those who maintained SR. This finding suggests that the presence of more organized and stable $f$-waves reduces the probability of long-term AF recurrence after SA. To this respect, more organized $f$-waves have been associated with less heterogeneous atrial electrical conduction during the arrhythmia, and therefore with less altered arrhythmogenic atrial substrate [23]. In view of these results and observations, it could also be considered that the analysis of how the $f$-waves evolve over time could provide more information about the patient's proarrhythmic condition after SA and the associated post-surgery treatments than other indices computed from long ECG intervals, where atrial information is somewhat averaged, such as DAF and FWA. Thus, this work opens a new door to gain new insights about the atrial substrate remodeling, which merits to be carefully explored in the future.

## 5. Conclusions

A novel methodology to predict preoperatively AF recurrence a year after surgical ablation has been introduced in the present work. The method was based on estimat-ing time variability of the fibrillatory waves energy, computed through a stationary wavelet decomposition, and has substantially improved the performance of other well-established parameters, such as DAF and FWA. As a consequence, the short-time analysis of how the fibrillatory waves evolve over time seems to provide new insights about the patient's proarrhythmic condition after the surgical procedure. This information could be useful to rationalize the management of potential candidates for the treatment, and hence opens an interesting research line.

## Acknowledgment

This research has been supported by grants DPI2017-83952-C3 from MINECO/AEI/FEDER EU, SBPLY/17/180501/000411 from Junta de Comunidades de Castilla-la Mancha and AICO/2019/036 from Generalitat Valenciana. Moreover, Pilar Escribano holds a graduate research scholarship from University of Castilla-La Mancha.

Address for correspondence:

Juan Ródenas García
E.S.I. Informática, Campus Univ., 02071, Albacete, Spain
Phone: +34-967-599-200 Ext. 2556
e-mail: juan.rodenas@uclm.es